\documentclass[pdflatex,sn-mathphys-num]{sn-jnl}
\usepackage{xr}
\usepackage[T1]{fontenc}
\usepackage{lmodern}
\usepackage{amsmath,amssymb,bm,booktabs}
\usepackage{microtype}
\usepackage{graphicx}
\usepackage{xcolor}
\usepackage{array}
\usepackage{siunitx}
\usepackage{multirow}
\usepackage{textcomp}
\usepackage{booktabs}
\usepackage{hyperref}

\newcommand{\ii}{\mathrm{i}}

\newcommand{\dd}{\mathrm{d}}

\newcommand{\ket}[1]{\lvert #1\rangle}
\newcommand{\bra}[1]{\langle #1\rvert}

\newcommand{\ee}{\mathrm{e}}

\newcommand{\RC}{\mathcal{R}_C}
\newcommand{\RT}{\mathcal{R}_M}
\newcommand{\RTk}{\mathcal{R}_{M\boldsymbol{k}}}
\newcommand{\RCT}{\mathcal{R}_{CM}}

\hypersetup{pdfauthor={},pdfsubject={Orbital magnetism and electron-scattering identifiability}}

\title{Ultrafast imaging of ring currents in atoms and condensed matter by electron scattering}

\author[1,2]{Zhongchen Xing}
\author[3]{Xinbao Liu}
\author[4,5]{Peixiang Lu}
\author[3,6]{Eberhard K. U. Gross*}
\email{eberhard.gross@mail.huji.ac.il}
\author[4,5]{Zheng Li*}
\email{zhengli@wit.edu.cn}

\affil[1]{State Key Laboratory for Mesoscopic Physics and Collaborative Innovation Center of Quantum Matter, School of Physics, Peking University, Beijing 100871, China}
\affil[2]{Collaborative Innovation Center of Extreme Optics, Shanxi University, Taiyuan, Shanxi 030006, China}
\affil[3]{Tsientang Institute for Advanced Study, Xihu District, 310024 Hangzhou, Zhejiang, China.}
\affil[4]{School of Science and Hubei Key Laboratory of Optical Information and Pattern Recognition, Wuhan Institute of Technology, Wuhan 430205, China}
\affil[5]{Wuhan National Laboratory for Optoelectronics and School of Physics, Huazhong University of Science and Technology, Wuhan 430074, China}
\affil[6]{Fritz Haber Center for Molecular Dynamics, Institute of Chemistry, The Hebrew University of Jerusalem, Jerusalem 91904 Israel}

\date{\today}
\begin{document}
\maketitle

\begin{abstract}

 Ring currents are well known to generate magnetic fields. Despite the rich phenomena and detection of the ring current, an experimental reconstruction of the ring current at the atomic scale is still unsolved. In this work, we propose a method of resolving the ring current in a sample at the atomic scale by ultrafast electron diffraction. Inspired by the Rosenbluth formula \cite{Rosenbluth1950}, we study the electron diffraction of a sample with a ring current and separate the scattering cross sections into three kinds of response functions, which stand for the electric part, the magnetic part and the cross term, respectively. The ring current in the sample can be reconstructed from measuring the response function in different scattering configurations. Also, following the Rosenbluth formula, we propose that the response functions can be measured by fitting over fixed momentum transfer with different incident energy and relative direction, which is similar to the $\psi$-scan in crystal diffraction. With high-precision ultrafast electron diffraction, the magnetic part will be resolved despite being much weaker than the electric part. We simulate three examples with ring currents -- an ideal hydrogen atom, a circularly polarized helium atom ensemble, and a lithium crystal. Our model offers a promising way to resolve the current structure at the atomic scale.
\end{abstract}

The question of the internal structure of atoms, molecules, and condensed matter by electron diffraction and electron scattering is one of the most important topics in physics. One of the most famous discoveries by electron scattering is the electromagnetic form factors of the proton. The definition and measurement are based on the Rosenbluth formula, first derived by Rosenbluth in 1950 \cite{Rosenbluth1950}. It defines the electromagnetic form factors in the relativistic quantum mechanics formalism, and provides a fundamental connection between elastic electron scattering cross sections and the form factors. Within the one-photon-exchange regime, the unpolarized differential cross section can be expressed in terms of two independent Sachs form factors, $G_E(Q^2)$ and $G_M(Q^2)$, which characterize the charge and current structures of the proton, respectively \cite{EICHMANN2016,Chen2023}. Besides defining the form factors, another inspiring point of the Rosenbluth formula is the extraction of the electromagnetic form factors. The total cross section is a summation of $G_E(Q^2)$ and $G_M(Q^2)$ with different dynamical weights. A particularly important consequence is that, at fixed four-momentum transfer $Q^2$, the reduced cross section depends linearly on the virtual-photon polarization parameter $\epsilon$, allowing the squared electric and magnetic form factors to be extracted separately through the Rosenbluth separation. This method established a cornerstone of experimental investigations of nucleon electromagnetic structure and remains an important benchmark for form-factor measurements \cite{Qattan2005}.

At the atomic scale, electromagnetic structure is a composite of the nuclei and the bound electrons. The bound electrons offer a current structure at the atomic scale via the ring current \cite{Barth2006,BarthManz2007}. Electronic ring currents, characterized by the circulating motion of electronic charge, represent a fundamental quantum phenomenon with far-reaching implications across atomic, molecular, and condensed-matter physics. 
% By generating orbital magnetic moments and local magnetic fields, these currents provide valuable insight into electronic angular momentum, quantum coherence, and charge delocalization beyond what can be inferred from the electron density alone.
In atomic systems, the preparation and detection of ring currents have enabled the investigation of electronic dynamics on attosecond timescales \cite{Eckart2018}.
In molecular systems, magnetically induced ring currents play a central role in understanding aromaticity and antiaromaticity, providing experimentally accessible signatures of electronic delocalization through nuclear magnetic resonance spectroscopy and magnetic response measurements \cite{Renana2015,Peeks2017}.
In condensed matter, persistent currents in mesoscopic rings reveal the macroscopic manifestation of quantum phase coherence \cite{Jayich2009}, while circulating electronic currents contribute to orbital magnetization and are involved in unconventional magnetic ordering in correlated materials \cite{Thonhause2005,Fernandes2026}.
Beyond these equilibrium phenomena, advances in ultrafast laser technology have stimulated considerable interest in generating and controlling transient electronic ring currents using tailored optical fields \cite{Katsuyuki2007,Tennesse2023}, opening new possibilities for manipulating microscopic magnetic fields and probing coherent electronic motion. Together, these developments establish ring currents not only as a fundamental manifestation of quantum electronic dynamics but also as a promising platform for investigating and controlling magnetic phenomena across a broad range of physical systems.

Measurements for electronic ring currents have been proposed and established across atomic, molecular, and condensed-matter systems. In argon atoms, attosecond-sensitive evidence of the preparation and detection of atomic ring currents is revealed by measuring strong-field ionization \cite{Eckart2018}. A ring current-inducing magnetic field is measured by nuclear magnetic resonance spectroscopy at \SI{7.5}{nm} level \cite{Peeks2017}. In aromatic molecules, time-resolved circular dichroism (TRCD) in the X-ray regime is proposed for resolving the magnitude and direction \cite{Yeonsig2020}. Despite these advances, the direct reconstruction of the electronic current-density distributions at the atomic scale remains a major experimental challenge. Developing current-sensitive probes capable of resolving both the spatial distribution and temporal evolution of electronic circulation is therefore an important direction for future research.

Electron scattering has been one of the most powerful experimental tools for investigating the microscopic structure and dynamics of matter since the pioneering electron diffraction experiments of Davisson and Germer \cite{Davisson1927}. Owing to their short de-Broglie wavelengths and strong electromagnetic interactions with charged constituents, electron beams provide exceptional sensitivity to the spatial distribution from the atomic scale to nuclei scale. In elastic scattering, the angular distribution of the scattered electrons encodes structural information through the Fourier components of the target's electrostatic potential, whereas inelastic scattering provides access to electronic excitations, molecular vibrations, phonons, plasmons, and other collective modes  \cite{Wolf2022,Kepaptsoglou2025}. Later, advances in ultrafast electron sources and measurements opened the gate to probing ultrafast processes. In particular, ultrafast electron diffraction (UED) has revealed photoinduced structural phase transitions \cite{Morrison2014} and probed exciton dynamics \cite{Kurtz2024} in solids, coherent nuclear wave-packet motion and nonadiabatic molecular dynamics near conical intersections \cite{Yang2018}, quantum state tomography of molecules \cite{Zhang2021,Zhang2022,Jiang2024}, and the simultaneous evolution of electronic and nuclear degrees of freedom in photoexcited molecules \cite{Yang2020}.
% Meanwhile, momentum-resolved electron energy-loss spectroscopy (EELS) has enabled nanoscale measurements of phonon dispersion [7], and recent time- and momentum-resolved experiments have uncovered the interplay between electronic excitations, phonons, and plasmons in nonequilibrium materials [8].
These developments span a broad range of incident electron velocities, from the nonrelativistic regime to highly relativistic MeV beams, and probe momentum transfers ranging from well below to above the characteristic inverse atomic length scale, namely the inverse of the Bohr radius \(a_0^{-1}\). Beyond conventional structural imaging, small-angle electron scattering offers opportunities to investigate electronic correlations and quantum fluctuations of the Coulomb potential \cite{Chen2024}. Looking forward, improvements in the simultaneous resolution of energy, momentum, time, and scattering intensity \cite{Diaz2024}, together with advanced theoretical descriptions of elastic and inelastic scattering, may enable increasingly precise measurements.

On the atomic scale, it is well known that circularly polarized laser pulses can induce electron ring currents in atomic and molecular systems\cite{Barth2006,Heras2023}. The ring currents then generate strong magnetic fields, as well as non-zero magnetic dipole moment \cite{BarthManz2007,Renana2015}. However, despite the symmetry breaking in many physical processes by the ring currents \cite{Eckart2018,Bredtmann2014}, the direct measurement of the magnetic field of the ring currents is less discussed.

We believe that ultrafast electron diffraction is a promising way to reconstruct ring currents at the atomic scale. In this work, we propose a method to extract information beyond the electron charge density within a quantum target by ultrafast electron scattering. The information about the magnetic field generated by the ring currents within the system is probed. The method is inspired by the Rosenbluth formula, which is used in extracting the electric and magnetic form factors by fitting against different scattering angles.

\section*{Relativistic coupling and the scattering observable}
\label{sec:dirac-main}

\begin{figure}[h]
    \centering
    \includegraphics[width=0.8\linewidth]{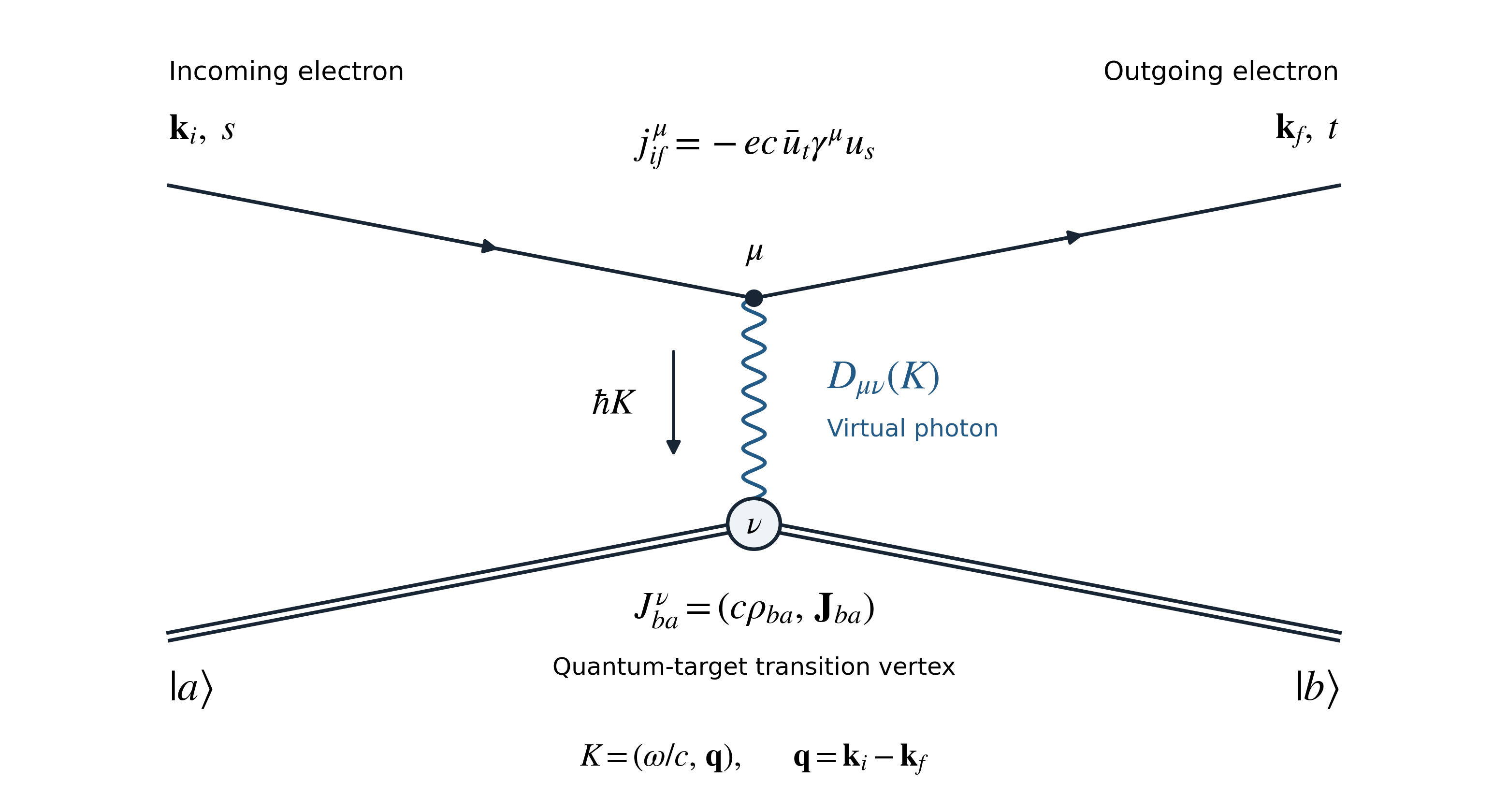}
    \caption{The lowest order Feynman diagram of $e^-+a\to e^-+b$. The upper solid line denotes the projectile electron, with incoming momentum/spin $(\boldsymbol k_i,s)$ and outgoing momentum/spin $(\boldsymbol k_f,t)$; its arrows show electron-line orientation. The upper vertex carries $j_{p,ts}^{\mu}=-ec\bar u_t(\boldsymbol k_f)\gamma^\mu u_s(\boldsymbol k_i)$. The lower double lines denote the initial and final many-body target states $|a\rangle$ and $|b\rangle$, and the open vertex represents their physical current matrix element $J_{ba}^{\nu}=(c\rho_{ba},\boldsymbol J_{ba})$, rather than an elementary-particle vertex. The wavy line is the virtual-photon propagator $D_{\mu\nu}(K)$. The separate downward arrow defines four-momentum transfer $\hbar K$ from projectile to target, with $K=(\omega/c,\boldsymbol q)$, $\boldsymbol q=\boldsymbol k_i-\boldsymbol k_f$, and $\hbar\omega=E_b-E_a$.}
    \label{fig:Feynman_diagram}
\end{figure}

Our model is built within relativistic quantum mechanics. We start from the lowest-order connected Feynman diagram for $e^-+a\to e^-+b$, where $a$ and $b$ stand for the initial and final states of a quantum target, as shown in Fig.~\ref{fig:Feynman_diagram}
The quantum target can be an atomic ensemble or a condensed matter system. The direct Fourier interaction kernel of the Feynman diagram is 
\begin{equation}
    \begin{aligned}
        \mathcal V_{bt;as}
=\frac{1}{\epsilon_0c^2}\,
j_{if,ts}^{\mu}(K)D_{\mu\nu}(K)J_{ba}^{\nu}(-K),
\label{eq:F_kernel}
    \end{aligned}
\end{equation}
where $D(q)$ is the reduced time-ordered photon propagator and $K^\mu = k_i^\mu - k_f^\mu=(\omega/c,\boldsymbol{q})$ is the 4-momentum transfer.  The subscripts $t$ and $s$ stand for the spin of the incident and outgoing electron. The 4-current form factors $j_{p,ts}^{\mu}=-ec\bar u_t(\boldsymbol k_f)\gamma^\mu u_s(\boldsymbol k_i)=(c\rho_p,\boldsymbol j_p)$ and $J_{ba}^{\mu}=(c\rho_{ba},\boldsymbol J_{ba})$ are the two transition vertices of the diagram. To solve the scattering cross section from Eq.~\eqref{eq:F_kernel}, we choose the Coulomb gauge for the propagator. The Fourier kernel of the Feynman diagram yields
\begin{equation}
    \begin{aligned}
        \mathcal V_{ba}=
 \frac{\rho_{if}\rho_{ba}}{\epsilon_0q^2}
 -\frac{\boldsymbol{j} _{if,T}\cdot\boldsymbol{J} _{ba,T}}
 {\epsilon_0c^2(\boldsymbol{q}^2-\omega^2/c^2-\mathrm{i} 0)}.
    \end{aligned}
    \label{Kernal_atom}
\end{equation}
The subscript $T$ stands for the transverse component, where the transverse component of a vector $\boldsymbol{\Lambda}$ is perpendicular to $\boldsymbol{q}$, namely $\boldsymbol \Lambda_{T}=\boldsymbol \Lambda-\hat{\boldsymbol q}(\hat{\boldsymbol q}\cdot\Lambda)$.
We further define the transition scalar potential $\Phi_{ba}$ and the transition vector potential $\boldsymbol{\mathcal{A}}_{ba}$ as
\begin{equation}
    \begin{aligned}
        \Phi_{ba}=&\,\frac{\rho_{ba}}{\epsilon_0q^2}\\
        \boldsymbol{\mathcal A}_{ba}=&\,\frac{\boldsymbol J_{ba,T}}{\epsilon_0c^2D_q}.
    \end{aligned}
\end{equation}
The Fourier kernel then reads
\begin{equation}
    \begin{aligned}
        \mathcal V_{ba}=\rho_{if}\Phi_{ba}-\boldsymbol j_{if}\cdot\boldsymbol{\mathcal A}_{ba}\equiv \mathcal V_C+\mathcal V_T.
    \end{aligned}
    \label{F_kernel_potential}
\end{equation}
In Eq.~\eqref{F_kernel_potential}, $\mathcal V_C$, $\mathcal V_T$ denote the scalar Coulomb contribution and the complete transverse-current contribution.
$\mathcal V_C$ can be regarded as the electric part
On the other hand, $\mathcal V_T$ is the transverse-current interaction, which includes the magnetic field contribution and the transverse electric field contribution $\boldsymbol{E}_T = -\partial_t \boldsymbol{A}$. In a stationary elastic channel, the transverse electric-field contribution vanishes, and the contribution is regarded as a magnetic field. The transition vector potential originates from the transverse current form factor. So, for quantum targets with ring currents, like a circularly excited atom, its transition vector potential $\boldsymbol{\mathcal A}_{ba}$ is non-zero. Therefore, the scattering amplitude also carries information about the ring currents besides the charge density through the magnetic field. 

The scattering amplitude is defined as $f^D_{ba}=-\frac{\varepsilon_f}{2\pi\hbar^2c^2}\mathcal V_{ba}.$, 
and the scattering cross section reads,
\begin{equation}
    \begin{aligned}
\frac{\mathrm d\sigma_a}{\mathrm d\Omega}
=\sum_{b}
\frac{\varepsilon_i\varepsilon_f}{4\pi^2\hbar^4c^4}\frac{k_f}{k_i}\times\frac12\sum_{s,t}\left|
\frac{\rho_{if}^{ts}\rho_{ba}}{\epsilon_0q^2}
-\frac{\boldsymbol j_{if,T}^{ts}\cdot\boldsymbol J_{ba,T}}
{\epsilon_0c^2(q^2-\omega^2/c^2-\mathrm i0)}
\right|^2.
\end{aligned}
\label{eq:scattering_cross_section}
\end{equation}
Here we focus on a spin-unpolarized incident electron beam, so we sum over all different spin channels evenly and incoherently. The transition 4-current form factor between the plane wave states $j^\mu_{if}$ reads
\begin{equation}
    \begin{aligned}
\rho_{if}^{ts}
&=-eN_fN_i\chi_t^\dagger\left[
\left(1+\frac{c^2\hbar^2\boldsymbol k_f\cdot\boldsymbol k_i}
{(\varepsilon_f+m_ec^2)(\varepsilon_i+m_ec^2)}\right)I_2
+\frac{\mathrm i c^2\hbar^2\boldsymbol\sigma\cdot(\boldsymbol k_f\times\boldsymbol k_i)}
{(\varepsilon_f+m_ec^2)(\varepsilon_i+m_ec^2)}\right]\chi_s,\\
\boldsymbol j_{if}^{ts}
&=-ec^2\hbar N_fN_i\chi_t^\dagger\Bigg[
\left(\frac{\boldsymbol k_i}{\varepsilon_i+m_ec^2}
+\frac{\boldsymbol k_f}{\varepsilon_f+m_ec^2}\right)I_2+\mathrm i\left(\frac{\boldsymbol k_i}{\varepsilon_i+m_ec^2}
-\frac{\boldsymbol k_f}{\varepsilon_f+m_ec^2}\right)\times\boldsymbol\sigma
\Bigg]\chi_s,
\end{aligned}
\label{eq:plane_wave_transition_current}
\end{equation}
where $N_{i,f} = \sqrt{\frac{\varepsilon_{i,f} +m_ec^2}{2\varepsilon_{i,f}}}$ is the normalization factor. 

\begin{figure}
    \centering
    \includegraphics[width=0.8\linewidth]{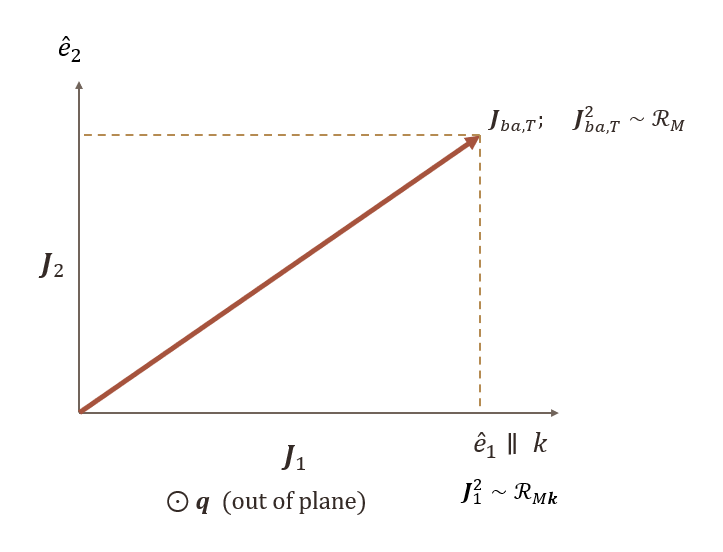}
    \caption{The relationship between the transverse response functions and the momenta. In elastic scattering, the momentum transfer $\boldsymbol{q}$ and the total momentum $\boldsymbol{k}$ have $\boldsymbol{q}\cdot \boldsymbol{k} = 0$. The transverse current form factor's parallel and orthogonal components against the total momentum contribute to $\mathcal{R}_{T\boldsymbol{k}}$ and $\mathcal{R}_T$, respectively. In atomic experiments, if the magnetic momentum of the ring current in real space is perpendicular to both $\boldsymbol{q}$ and $\boldsymbol{k}$, the cross term $\mathcal{R}_{CT}$ will be canceled.}
    \label{fig:R_T_and_R_Tk}
\end{figure}

Subsitude Eq.~\eqref{eq:plane_wave_transition_current} into Eq.~\eqref{eq:scattering_cross_section}. We further define the following response functions as
\begin{equation}
    \begin{aligned}
        \RC&=
        \frac{1}{e^2}|\rho_{ba}|^2,\\
\RTk&=\frac1{e^2c^2}
|\hat{\boldsymbol k}_T\cdot\boldsymbol J_{ba,T}|^2,\\
\RT&,\\
\RCT&=\frac{1}{e^2c}\operatorname{Re}[
\rho_{ba}^*\hat{\boldsymbol k}_T\cdot\boldsymbol J_{ba,T}].
    \end{aligned}
    \label{eq:response_function}
\end{equation}
And the corresponding dynamical weights,
\begin{equation}
    \begin{aligned}
v_C&=\frac{\varepsilon_i\varepsilon_f+m_e^2c^4+\hbar^2c^2\boldsymbol k_i\cdot\boldsymbol k_f}
{2\varepsilon_i\varepsilon_f},\\
v_{Mk}&=\frac{\hbar^2c^2q^4|\boldsymbol k_T|^2}{4\varepsilon_i\varepsilon_f\mathcal D^2},
\\v_M&=\frac{\hbar^2c^2q^4}{4\varepsilon_i\varepsilon_f\mathcal D},\\
v_{CM}&=-\frac{\hbar c(\varepsilon_i+\varepsilon_f)q^2|\boldsymbol k_T|}
{2\varepsilon_i\varepsilon_f\mathcal D},
\end{aligned}
\end{equation}
where $\mathcal{D} = q^2 - \omega^2/c^2 >0$, $\boldsymbol{q} = \boldsymbol{k}_i -\boldsymbol{k}_f$, and $\boldsymbol{k} = \boldsymbol{k}_i+\boldsymbol{k}_f$. The cross section can be reduced into the following form,
\begin{equation}
    \Sigma\equiv\frac{\epsilon_0^2q^4}{\mathcal N e^2}
\frac{\mathrm d\sigma}{\mathrm d\Omega\,}
=v_C\RC+v_{Mk}\RTk
+v_M\RT+v_{CM}\RCT.
\label{eq:reduced_cross_section}
\end{equation}
In Eq.~\eqref{eq:response_function} and in the elastic scattering case, the response function $\RT$ can be regarded as the $G_M^2$ in the Rosenbluth formula. The transverse current density $\boldsymbol J_{ba,T}(\boldsymbol q)\simeq-\mathrm i\,\boldsymbol q\times
\langle b|\hat{\boldsymbol\mu}(\boldsymbol{q})|a\rangle.
$ is orthogonal to the momentum transfer. In a hydrogen-like atom, the magnetic moment operator reads,
\begin{equation}
\hat{\boldsymbol\mu}=-\mu_B\frac{\hat{\boldsymbol L}}{\hbar},
\end{equation}
which comes directly from the ring current. The response function $\RTk$, on the other hand, is a new term compared to the Rosenbluth formula. The difference $\RT - \RTk$ represents the contribution from projecting $\boldsymbol\mu$ onto the scattering plane. If $\boldsymbol{\mu}$ is perpendicular to the scattering plane and $b=a$, then $\RTk = \RT$. The inequality between $\RT \neq \RTk$ turns out ot be the leakage to other states with $b\neq a$. An illustrative figure of the transverse functions, namely $\mathcal{R}_T$, $\mathcal{R}_{T\boldsymbol{k}}$, and $\mathcal{R}_{CT}$ is shown in Fig.~\ref{fig:R_T_and_R_Tk}.
For the elastic channel with $b=a$, the response function can be used in reconstructing the field structure up to a sign. Define
\begin{equation}
    Q = \rho_{aa}(\boldsymbol{q})/e,\quad \mathcal{J}_{1,2} = \frac{\hat{e}_{1,2}\cdot \boldsymbol{J}_{aa}(\boldsymbol{q})}{e}.
\end{equation}
Here $\hat{e}_{1,2}\perp \boldsymbol{q}$.
Since in elastic scattering $\boldsymbol{k} \perp \boldsymbol{q}$, we denote the total momentum $\boldsymbol{k}$ as 
\begin{equation}
    \boldsymbol{k} = k(\hat{e}_1\cos\phi +\hat{e}_2\sin\phi).
\end{equation}
We further introduce the following parameters
\begin{equation}
C=|Q|^2,\qquad
S_{ij}=\operatorname{Re}(\mathcal{J}_i^*\mathcal{J}_j),\qquad
h_i=\operatorname{Re}(Q^*\mathcal{J}_i).
\end{equation}

Then, the response functions have
\begin{equation}
\begin{aligned}
\RC &= C\\
\RTk(\psi)
&=S_{11}\cos^2\phi+S_{22}\sin^2\psi
+2S_{12}\sin\psi\cos\psi,\\
\RT&=S_{11}+S_{22},\\
\RCT(\psi)&=h_1\cos\psi+h_2\sin\psi.
\end{aligned}
\label{eq:measurement}
\end{equation}
All parameters $C,\,S_{11},\,S_{22},\,S_{12},\,h_1,\,\text{and }h_2$ are solvable with scanning $\psi$. Then, up to an undetermined phase, we can reconstruct the charge density and current density form factors as
\begin{equation}
Q=e^{i\chi}\sqrt C,\qquad
\boldsymbol{\mathcal{J}}=e^{i\chi}
\left(\frac{\boldsymbol h}{\sqrt C}\pm i\boldsymbol v\right),
\label{charge_current_reconstruction}
\end{equation}
where $\boldsymbol{v}$ has,
\begin{equation}
    S - \boldsymbol{hh^\mathrm{T}}/C = \boldsymbol{v}\boldsymbol{v}^{\mathrm{T}}.
\end{equation}
For ring current eigenstates of the hydrogen-like atoms, $\rho(\boldsymbol{q})$ is real and $\boldsymbol{J}(\boldsymbol{q})$ is imaginary, provided that other states do not participate in the elastic scattering cross section. Then $\mathcal{R}_{T\boldsymbol{k}} = \mathcal{R}_T$ and $\chi=0$ or $\pi$, which is similar to the reconstruction of proton structure by the Sache decomposition \cite{Chen2023}

%Please check
The proposed measurement can be implemented by adapting existing laser-pump/electron-probe diffraction techniques, such as ultrafast low-energy electron diffraction (ULEED) \cite{Kurtz2024} and ultrafast reflection high-energy electron diffraction (URHEED) \cite{Horn2024}. The sample is mounted on a multi-axis goniometer, allowing the momentum-transfer vector $\mathbf q=\mathbf k_f-\mathbf k_i$, which is determined by the detectors, to be aligned along a selected direction in the material frame. Subsequently, the sample is rotated about the scattering vector $\mathbf q$, while the electron source and detector remain stationary. This procedure, known as an azimuthal $\psi$-scan, is well established in single-crystal diffraction \cite{Busing1966}. Such a rotation leaves $\mathbf q$ unchanged in the material frame while continuously varying the direction of $\mathbf k=\mathbf k_i+\mathbf k_f$ in the plane perpendicular to $\mathbf q$. To maintain identical optical excitation conditions during the azimuthal scan, the pump-beam propagation direction and polarization must remain fixed in the material frame. This can be achieved using computer-controlled beam-steering mirrors and polarization optics, whose settings are adjusted according to the sample orientation. Similar motorized optical alignment systems have been implemented in time-resolved X-ray scattering \cite{Burian:ju5002}, while active pump-beam pointing and position stabilization have been demonstrated in ultrafast electron diffraction \cite{Li2022Duncan}. The experimental design is shown in 
\begin{figure}
    \centering
    \includegraphics[width=0.8\linewidth]{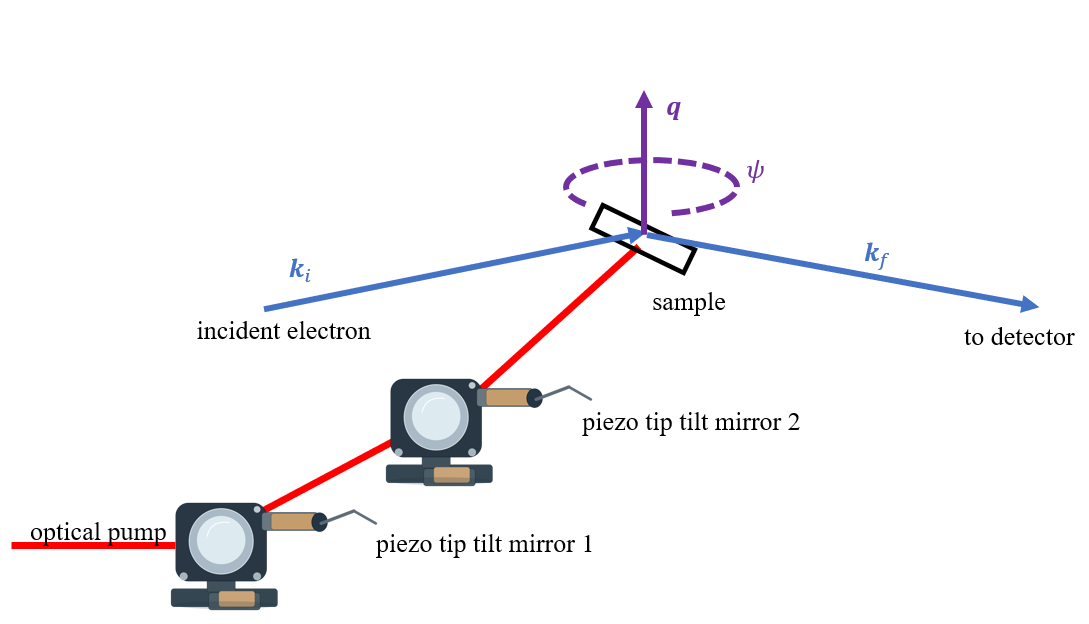}
    \caption{The demonstration of the reconstruction of ring current inside a sample. First, measure the position of the detector, which determines the momentum transfer $\boldsymbol{q}$. Next, choose the direction of $\boldsymbol{q}$ as the axis; both the sample and the optical pump are rotating about it. The rotating of the sample is also known as the $\psi$-scan, while the optical pump can be controlled by two piezo tip tilt mirrors. During the rotation, the incident electron beam and the detector are unchanged. To vary $\boldsymbol{k}$ while leaving $\boldsymbol{q}$ fixed, one can change the energy of the electron beam and adjust the detector properly. Combining those operations and Eq.~\eqref{14} leads to the reconstruction of the current density. }
    \label{fig:reconstruction_protocal}
\end{figure}

As an intuitive example, we study the magnetic form factor of the $\ket{a} =\ket{n=2,l=1,m=+1}$ state of a hydrogen atom. The charge density and the current density of it are \cite{BarthManz2007}
\begin{equation}
    \begin{aligned}
        n_{+1}(\boldsymbol r)&=\frac{r_x^2+r_y^2}{64\pi a_0^5}\ee^{-r/a_0},\\
\rho_{+1}(\boldsymbol r)&=e\delta^3(\boldsymbol r)-e n_{+1}(\boldsymbol r),\\
\boldsymbol j_{+1}(\boldsymbol r)
&=-\frac{e\hbar}{m_e}\operatorname{Im}(\psi_{+1}^*\nabla\psi_{+1})
=-\frac{e\hbar \ee^{-r/a_0}}{64\pi m_ea_0^5}\hat{z}\times \boldsymbol{r}.
    \end{aligned}
\end{equation}
The magnetic dipole momentum contributed by the excited hydrogen atom is $\mu_z = -\mu_B$ with $\mu_B = \frac{e\hbar}{2m_e}$.

For elastic scattering, the final state of the quantum target is a superposition of $\ket{2p,m=-1,0,1}$ and $\ket{2s}$ in general. Take
\begin{equation}
    \ket{P_z} = \ket{2p,0},\, \ket{P_x} = \frac{\ket{2p,-1}-\ket{2p,1}}{\sqrt{2}}, \, \ket{P_y} = \ii\frac{\ket{2p,1}+\ket{2p,-1}}{\sqrt{2}},
\end{equation}
which forms the Cartesian basis of the 2p state. The transition current $J_{bc}^\mu$ then reads,
\begin{equation}
    \begin{aligned}
\rho_{bc}(\boldsymbol q)&=e\left[
\left(1-\frac1{(1+x^2)^3}\right)\delta_{bc}
+\frac{6a_0^2q_bq_c}{(1+x^2)^4}\right],\\
\boldsymbol J_{bc}(\boldsymbol q)&=\frac{e\hbar}{2m_e(1+x^2)^3}
\left(q_c\hat{\boldsymbol e}_b-q_b\hat{\boldsymbol e}_c\right),\\
\rho_{2s,c}&=\frac{3e\mathrm i a_0q_c(1-x^2)}{(1+x^2)^4},\\
\boldsymbol J_{2s,c}&=\frac{\mathrm i e\hbar a_0}{2m_e(1+x^2)^3}
\left(q^2\hat{\boldsymbol e}_c-q_c\boldsymbol q\right),
\end{aligned}
\label{eq:2p_transition_current}
\end{equation}
where $b,c \in \{x,y,z\}$.

Substituting Eq.~\eqref{eq:2p_transition_current} into Eq.~\eqref{eq:response_function}, The response functions are
\begin{equation}
\begin{aligned}
\RC(x,c_q)
&=\frac{1-c_q^2}{2}
\left[1-\frac{1-5x^2}{(1+x^2)^4}\right]^2
+\frac{1+c_q^2}{2}
\left[1-\frac1{(1+x^2)^3}\right]^2
+\frac{9x^2(1-x^2)^2(1-c_q^2)}{2(1+x^2)^8},\\
\RTk(x,c_q,c_k)
&=\frac{d}{2(1+x^2)^6}
\left[2+x^2-c_q^2-(1+x^2)c_k^2\right],\\
\RT(x,c_q)
&=\frac{d}{2(1+x^2)^6}
\left[3+x^2+(x^2-1)c_q^2\right],\\
\RCT(x,c_q,c_k)
&=\frac{3\sqrt d\,x^2}{2(1+x^2)^6}\,c_qc_k,
\end{aligned}
\label{eq:H_explicit_response_functions}
\end{equation}
where $x=qa_0$ is the dimensionless momentum transfer, $d = \alpha_{\rm fs}^2x^2/4$ with $\alpha_{\rm fs}$ being the fine structure constant, and $c_q=\hat{\boldsymbol q}\cdot\hat{\boldsymbol z}$ and $c_k=\hat{\boldsymbol k}\cdot\hat{\boldsymbol z}$ are the direction cosines

\begin{figure}
    \centering
    \includegraphics[width=0.7\linewidth]{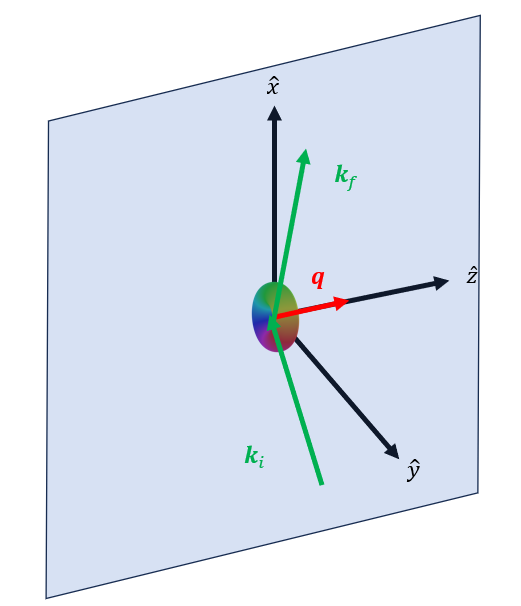}
    \caption{An illustrative figure of the experimental design for extracting the response functions. The system is polarized by circularly polarized light, inducing ring currents and a magnetic dipole moment mainly on the $z$-axis. The momentum transfer of the scattered electron lines is also along the $z$-axis. For elastic scattering, the total momentum $\boldsymbol{k}$ is perpendicular to $\boldsymbol{q}$. }
    \label{fig:H_protocal}
\end{figure}

Consider that the incident beam is perpendicular to the target $z$ axis and select the outgoing directions in the fixed lab $xy$ plane. Then $c_q^2 = 1 $ and $c_k^2=0$. Then, $\RTk = \frac{1}{2}\RT$ while $\RCT = 0$. $\mathcal{R}_M \neq \mathcal{R}_{M\boldsymbol{k}}$ is because the 2p states and 2s state are degenerate. The reduced cross section becomes,
\begin{equation}
    \begin{aligned}
\Sigma&=\frac{x^4}{4\gamma^2a_0^2}\frac{\mathrm d\sigma_{+1}}{\mathrm d\Omega}
=v_C\mathcal R_C+v_{M}^\prime\mathcal R_T,\\
v_C&=1-d+dt,\\
v_{M}^\prime&=\frac{1}{2}[t+d(1-t)],
\end{aligned}
\end{equation}
where $t = \beta^2 = v_i^2/c^2$
Then,
\begin{equation}
    \frac{\Sigma(\Omega)}{v_{T}^\prime}
=\frac{v_C}{v_{T}^\prime}\RC+\RT.
\label{eq:Rosenbluth_like}
\end{equation}
To measure the response functions in general with Eq.~\ref{eq:measurement}, one needs to fix $\mathcal{q}$, while rotating $\boldsymbol{k}$ on the transverse plane and varying the incident energy. A demonstration of the experimental setup is in Fig.~\ref{fig:H_protocal}

\begin{figure}[h]
    \centering
    \includegraphics[width=0.8\linewidth]{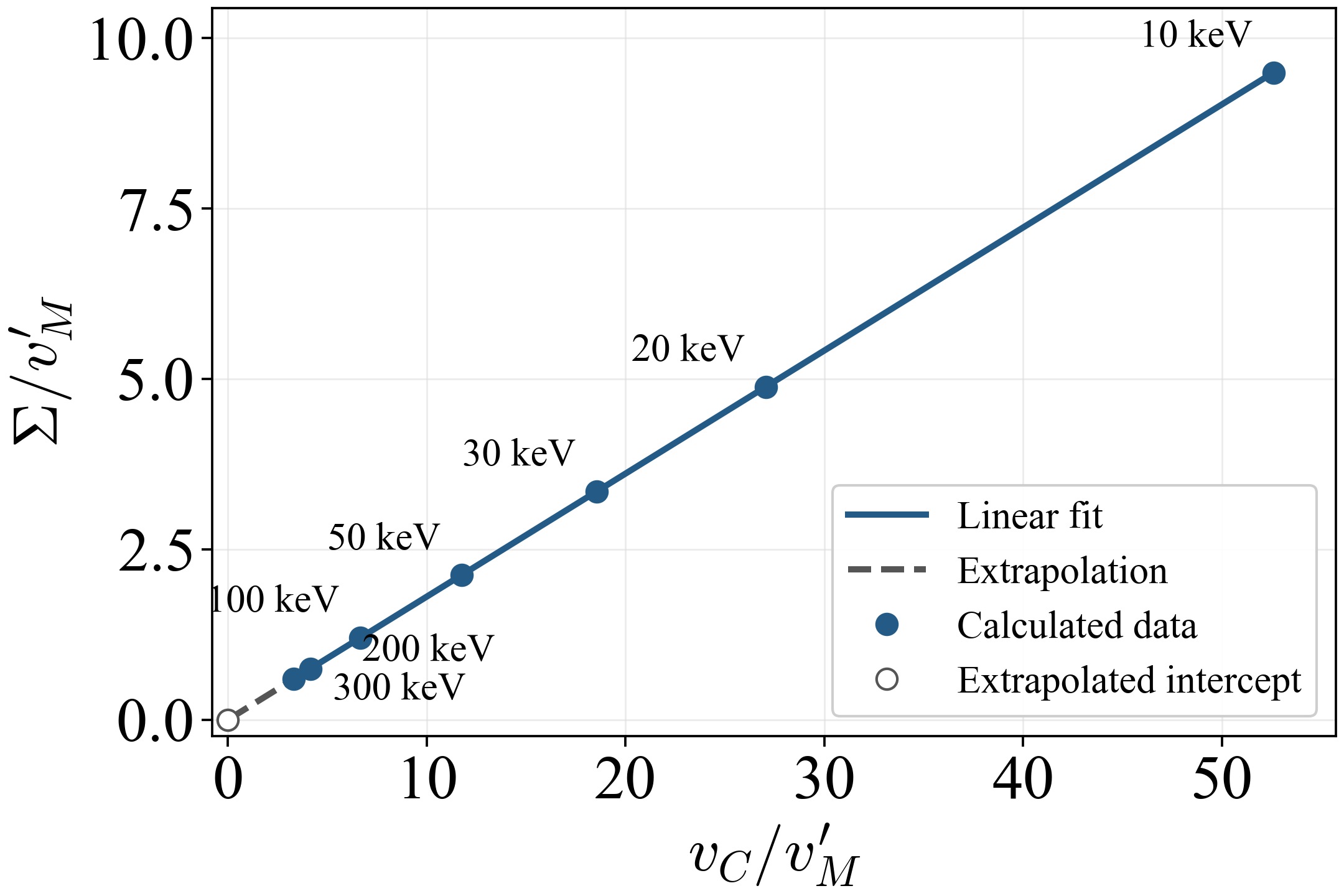}
    \caption{Extraction of $\RT$ of $\ket{2p,m=1}$ state of the hydrogen atom following Eq.~\eqref{eq:Rosenbluth_like}. The atom is assumed to be polarized along the $z$-axis. The elastic scattering in the $xz$ plane with $\boldsymbol{q} = q\hat{z}$ is collected. At fixed momentum transfer $qa_0=0.45$, kinetic energies 10, 20, 30, 50, 100, 200, and 300 keV are simulated.  The segment and star mark extrapolation to $X=0$, which represents the $\mathcal{R}_M$. In this example, $\mathcal{R}_C=0.181$ and $\mathcal{R}_M = 1.07\times 10^{-6}$, which requires a resolution at $10^{-6}$ precision.}
    \label{fig:H_example}
\end{figure}

So, by varying the incident energy while keeping $q$ unchanged, we will get a straight line against $v_C/v_{M}^\prime$, where its slope is $\RC$, and the intercept is the full transverse strength $\RT$, which has the orbital magnetic interpretation in this elastic model. A simulation of extracting $\RCT$ is shown in Fig.~\ref{fig:H_example}.

\section*{Magnetic Form factors in helium atom and lithium crystals}
We also applied our method to the He atom and the Li crystal.

The He atom is modeled as a six-correlated singlet model, containing three P sublevels, the ground state, and the metastable singlet state. The free Hamiltonian of the helium atom is
\begin{equation}
    H_0 = E_P\sum_{b=x,y,z}\ket{P_b}\bra{P_b} +E_g\ket{g}\bra{g}+E_s\ket{s}\bra{s},
\end{equation}
where $\ket{g}$, $\ket{s}$, and $\ket{P_\mu}$ are the ground state, the metastable singlet state, and the three orbital sublevels. The initial density matrix of the helium is set as $\rho_0 = 0.9\ket{g}\bra{g}+0.1\ket{s}\bra{s}$. We use a circularly polarized Gaussian pump that propagates along a fixed laboratory $z$ to prepare atoms with ring current. The pump is resonant with the metastable-S–P transition.

\begin{figure}
    \centering
    \includegraphics[width=1.0\linewidth]{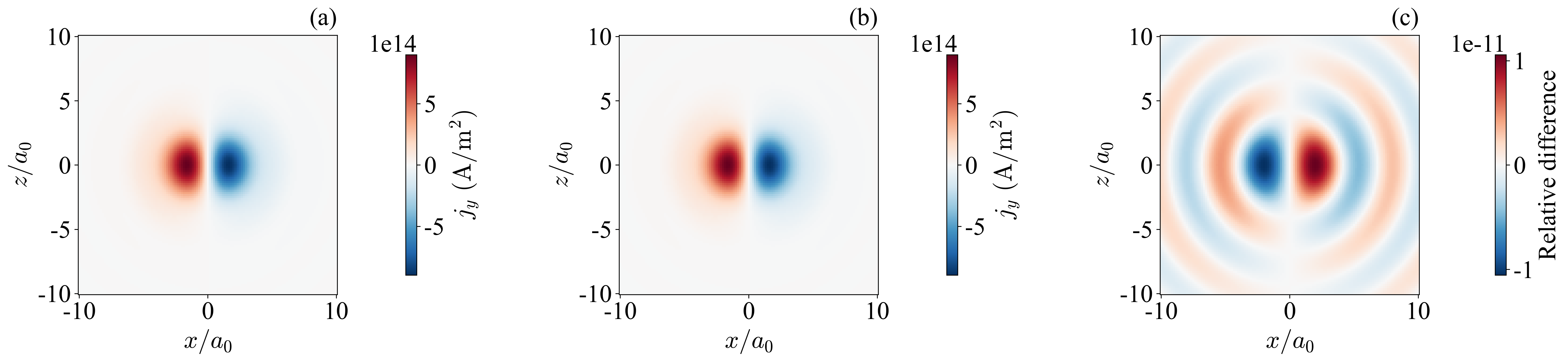}
    \caption{The reconstruction of ring current in helium by measuring the response functions at $y=0$ plane. Here, (a) is the simulated ring current, (b) is the reconstruction from the response functions, and (c) is the difference between (a) and (b). The calculation shows that $\chi=0$ assumption is good for the helium atoms. This is because the helium atom is spherically symmetric before optical pumping.}
    \label{fig:he_reconstruction}
\end{figure}

Here we simulate the ring current reconstruction of the helium atom. We calculated the response functions of the excited helium atom at $0.11\le qa_0\le2$. Choosed $32\times9\times8=2304$ grind points of $(q,\theta_q,\phi_q)$ in the lab frame. The ring current is reconstructed following Eq.~\ref{charge_current_reconstruction}, where the undetermined phase is assumed to be $\chi=0$. The reconstruction of helium is shown in Fig.~\ref{fig:he_reconstruction}.

% Same as the example of the hydrogen atom, we use $c_q^2=1$ and $c_k^2=0$, which leads to $\mathcal{R}_{CT} = 0$. The simulation on the helium ensemble is shown in Fig. \ref{fig:He_example}.
% \begin{figure}
%     \centering
%     \includegraphics[width=1.0\linewidth]{figures/helium_arrangement2_x045.png}
%     \caption{ The extraction of the magnetic form factor of a helium ensemble pumped by an ultrafast circularly polarized light. The momentum transfer is fixed at $x=qa_0=0.45$, with different electron beam energies. The pump's wavelength is \SI{2058.8}{nm} and its temporal FWHM is \SI{27.8}{ps}. The extraction yields $\mathcal{R}_C = 0.275$ and $\mathcal{R}_T=2.36\times 10^{-7}$. }
%     \label{fig:He_example}
% \end{figure}

We also applied our model to a Li crystal sample. We use electron scattering at different energies to probe the response functions of a circularly pumped Li crystal.  

The Li crystal is simulated in explicit memory. First-principles real-time implementation for nonlinear dynamics, namely
\begin{equation}
    \ii \frac{\partial \psi_j(\boldsymbol{r},t)}{\partial t} = \left\{ \frac{1}{2}\left[ -\ii \nabla +\boldsymbol{A}(\boldsymbol{r},t) +\boldsymbol{A}_{\rm xc}(\boldsymbol{r,t}) \right]^2 +v(\boldsymbol{r},t)+v_H(\boldsymbol{r},t)+v_{\rm xc}(\boldsymbol{r},t)  \right\} \psi_j(\boldsymbol{r},t)
    \label{eq:TDDFT}
\end{equation}

\begin{figure}
    \centering
    \includegraphics[width=1.0\linewidth]{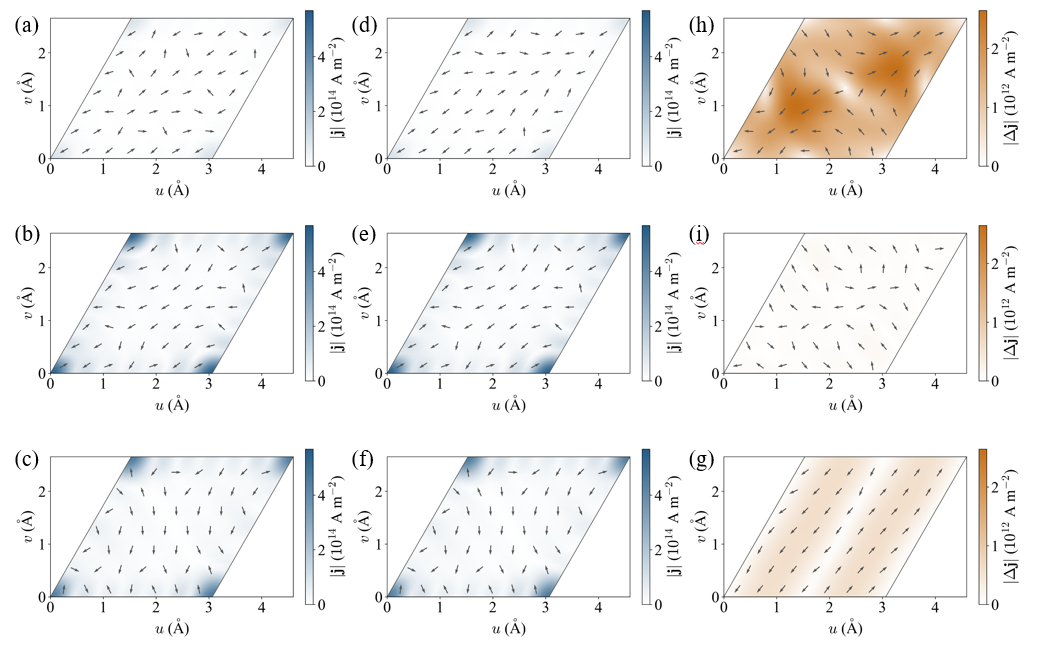}
    \caption{The reconstruction of the current inside lithium at $\frac{1}{2}\vec{a}_2$ plane. Here (a)-(c) are the simulated current at time delays $\tau = \SI{19.1}{fs}$, $\SI{19.4}{fs}$, and $\SI{19.7}{fs}$, respectively. Figures (d)-(f) are the reconstructed ring current, corresponding to (a)-(c). Figures (g)-(i) are the difference between the current and the reconstruction. Unlike in a gaseous medium, $q$ is determined by the reciprocal lattice vector. We use the response functions at $\boldsymbol{q}\sim(n_1n_2n_3)$, with $-3\le n_{1,2,3}\le 3$. $\boldsymbol{q}=0$ is excluded. The maximum error ratio between the current and the reconstruction is 3.86\%, which is caused by the undetermined phase $\chi$.}
    \label{fig:li_reconstruction}
\end{figure}

The correlation term follows
\begin{equation}
    \frac{\dd}{\dd t}\boldsymbol{A}_{xc,k}(\boldsymbol{r},t) = \frac{1}{n(\boldsymbol{r},t)}\nabla_i \sigma_{ik}(\boldsymbol{r},t).
\end{equation}

\begin{equation}
    \sigma_{ik}(\boldsymbol{r},t) = \int_{-\infty}^t \dd t^\prime \left\{
    \eta(\boldsymbol{r},t,t^\prime)[\nabla_{k}u_i(\boldsymbol{r},t)+\nabla_i u_k(\boldsymbol{r},t^\prime)-\frac{2}{3}\nabla\cdot \boldsymbol{u}(\boldsymbol{r},t)\delta_{ij}]+\zeta(\boldsymbol{r},t,t^\prime)\nabla\cdot \boldsymbol{u}(\boldsymbol{r},t^\prime)\delta_{ij}
    \right\}.
\end{equation}
And,
\begin{equation}
    f_{\rm xc}(n,\omega) = f_{\rm xc}(n,\infty)+\frac{1}{2}\sum_{m=1}^{M}\left[
    \frac{C_m(n)p_m(n)}{\omega - p_m(n)} - \frac{C^*_m(n)p_m*(n)}{\omega + p_m^*(n)}
    \right].
\end{equation}
The optical pump of the lithium is based on Eq.~\eqref{eq:TDDFT}. An ultrafast optical pump induces ring currents in the crystal and the corresponding magnetic field. We simulate the reconstruction of the ring current by ultrafast electron diffraction. The response functions are measured after the optical pump at the time delay of $\tau$.
% The magnetic field inside the primitive cell is shown in Fig.~\ref{fig:li_generated_field}.
% \begin{figure}
%     \centering
%     \includegraphics[width=0.8\linewidth]{figures/li_current_generated_field.png}
%     \caption{The magnetic field generated in a lithium crystal pumped by an ultrafast light at $\frac{1}{2}\vec{a}_2$ plane.}
%     \label{fig:li_generated_field}
% \end{figure}
The undetermined phase is also set as $\chi=0$. The reconstruction is shown in Fig.~\ref{fig:li_reconstruction}.

Our findings offer a method towards resolving the ring current at the atomic scale. By optical-electron pump-probe, the $\psi$-scan, and the Rosenbluth formula-like separation, the response functions can be determined. In the elastic scattering regime, one can reconstruct the ring current in inverse space up to a phase function $\chi{\boldsymbol{q}}$. We expect that our findings can be extended to other molecules and solids, and offer a more detailed description of the ring current.

\subsection*{Data and code availability}
The data and code are available upon reasonable request from the corresponding author.

\subsection*{Author Contributions}
Z.X. and X.L. performed the analysis and numerical calculations. Z.L. and E.K.U.G. contributed to the theoretical results. All authors contributed to the manuscript.

%\clearpage
%\begingroup
%\raggedright
%\bibliographystyle{unsrt}
\bibstyle{nature}
\bibliography{references}
%\endgroup
\end{document}